\documentclass[a4paper,11pt]{article}
\usepackage[utf8x]{inputenc}
\usepackage[bookmarks,breaklinks, colorlinks, urlcolor=magenta, citecolor=blue, link color=green]{hyperref}

\usepackage{amsmath,amssymb,graphicx,bm}
\usepackage{slashed}
\usepackage{color}
\usepackage[bottom]{footmisc}
\newcommand{\beq}{\begin{equation}}
\newcommand{\eeq}{\end{equation}}
\newcommand{\bea}{\begin{eqnarray}}
\newcommand{\eea}{\end{eqnarray}}
\newcommand{\ba}{\begin{align}}
\newcommand{\ea}{\end{align}}
\newcommand{\bfig}{\begin{figure}}
\newcommand{\efig}{\end{figure}}

\def\be {\begin{equation}}
\def\ee {\end{equation}}
\def\bea {\begin{eqnarray}}
\def\eea {\end{eqnarray}}
\def\bc {\begin{center}}
\def\ec {\end{center}}
\def\nn {\nonumber}
\begin{document}

\title{Consistency tests of AMPCALCULATOR and chiral amplitudes in SU(3) Chiral
Perturbation Theory:  A tutorial
based approach}

\author{B.\ Ananthanarayan$^a$, Diganta Das$^b$, I. Sentitemsu Imsong$^a$ \\
$^a$Centre for High Energy Physics, \\
Indian Institute of Science, Bangalore 560 012, India \\
$^b$The Institute of Mathematical Sciences \\
Taramani, Chennai 600113, India}

\maketitle

\begin{abstract}
Ampcalculator is a Mathematica$^{\textrm{\scriptsize\copyright}}$ based program that was made publicly
available some time ago by Unterdorfer and Ecker. It enables the user to 
compute several processes at one-loop (upto $O(p^4)$) in $SU(3)$ chiral 
perturbation theory. They include computing matrix elements and form factors for 
strong and non-leptonic weak processes with at most six external states. 
It was used to compute some novel processes and was tested against
well-known results by the original authors.  Here we present the results of
several thorough checks of the package. Exhaustive checks performed by the original authors 
are not publicly available, and hence the present effort.
Some new results are obtained from the software especially
in the kaon odd-intrinsic parity non-leptonic decay sector involving the coupling $G_{27}$.  
Another illustrative set of amplitudes at tree level we provide is in the context of
$\tau$-decays with several mesons including quark mass effects, of use to the BELLE experiment.
All eight meson-meson scattering amplitudes have been checked.
Kaon-Compton amplitude has been checked and a minor error in published results has
been pointed out. This exercise is a tutorial based one, wherein several input and output notebooks are also being
made available as ancillary files on the arXiv. Some of the additional notebooks we
provide contain explicit expressions that we have used for comparison with established results.
The purpose is to encourage users to apply the software
to suit their specific needs. An automatic amplitude
generator of this type can provide error-free outputs that could be used as inputs for
further simplification, and used in varied scenarios such as applications of chiral
perturbation theory at finite temperature, density and volume. This can also be used by students
as a learning aid in low-energy hadron dynamics.

\end{abstract}

\bigskip

\section{Introduction \label{sec:intro}}

$SU(3)$ chiral perturbation theory (ChPT) is a mature subject and has been over the
years tested in great detail.  Since the pioneering work
of Gasser and Leutwyler \cite{Gasser:1984gg, Gasser:1984ux},
many teams have worked hard and have produced a large body
of work and have computed processes of interest to experiment
and theory. The processes that have been computed include
form factors and scattering amplitudes of importance to electromagnetic
interactions and weak interactions. There are also odd-intrinsic parity
processes which have been computed.  Several non-leptonic decays
of kaons have also been studied which deals with a near independent
sector \cite{Cirigliano:2011ny}. In general, results have appeared in the literature over
the last couple of decades and virtually all processes that
are tractable and of interest to phenomenology and experiment
are now exhausted.

Some time ago, a very useful Mathematica$^{\textrm{\scriptsize\copyright}}$ based program that can
compute amplitudes in $SU(3)$ ChPT in the even-intrinsic parity and odd-intrinsic parity
(anomaly mediated) processes due to Unterdorfer and Ecker
(UE)\cite{Unterdorfer:2005au} has been
made publicly available.  With the exception of certain
anomalous processes, the program is capable of producing a
representation for form factors and scattering amplitudes
in the theory with user supplied input for the choice of particles
and momenta for up to six external particles (with a photon and
$W$-boson counting for 2 particles each). UE have developed the program
for evaluating amplitudes for some hitherto unstudied processes and also
to check amplitudes for known processes such as $e^+e^- \rightarrow 4 \pi$ 
and a $\tau \rightarrow 4\pi$ ~\cite{Ecker:2002cw}, of importance to, 
\emph{e.g}, TAUOLA~\cite{Jadach:1993hs,Shekhovtsova:2012ra}. More recently,
Ampcalculator (AMPC) was used to look at the decay $K\to \pi l^+ l^-$  \cite{AnantAMPC:2012} and
a missing $G_{27}$ piece in the SU(3) one-loop amplitudes was found.

In the light of the above, it is perhaps a useful exercise to
employ AMPC to try and build an exhaustive library of Mathematica$^{\textrm{\scriptsize\copyright}}$
based programs that can check the existing results in the literature
and alternatively to use the published results to check the consistency of AMPC.
Our aim is to provide a first attempt at such a compilation.
In many cases, we also provide Mathematica$^{\textrm{\scriptsize\copyright}}$ input
for each of the programs and the corresponding output notebooks obtained
by us.  It may be noted that AMPC may be sensitive to version
of Mathematica$^{\textrm{\scriptsize\copyright}}$ used, as it was first written in 
Mathematica$^{\textrm{\scriptsize\copyright}}$ 5.
Here we also provide a dictionary for translating the loop functions
coming out of AMPC denoted by the $A$ and $B$ into more familiar
functions. We also carry out some simple
tree-level computations which are of importance to $\tau$-decays.
Although the issue of the neglect of the quark masses was noted in
Finkemeier et al.\cite{Decker:1993ay}, even today experiments appear to use the work of
Aubrecht, et.al \cite{Aubrecht:1981cr}, especially when
$\eta$ mesons are in the final state \cite{Usuki:2009zza}.  In order to
draw the attention of the community to this, we carry out tree-level computations of all the
relevant processes using AMPC and provide a detailed comparison with the results of Ref.~\cite{Aubrecht:1981cr}, so that
experimentalists may update their data bases using information that does not neglect quark
masses. It is our belief that AMPC can provide readily accessible 
results also of importance to experimental efforts such as the BELLE.

The motivation of the present work is also to present a thorough comparison to
the extent possible with amplitudes and form factors that are sufficiently simple.  Amplitudes
that involve a large number of particles gives rise to results that are not easily amenable to
comparison.  Examples include $K\rightarrow 3 \pi$, $\tau\rightarrow 3 \pi \nu_\tau$ 
decays. We do not provide a comparison with these amplitudes. However, it should be noted that as recently as two years
ago, one of the AMPC accessible processes was computed in a heroic effort by Kaiser
\cite{Kaiser:2010zg} who computed the amplitudes for the processes $\pi^-\gamma\rightarrow3\pi$ diagram by
diagram. In future, AMPC could be employed for such practical needs.

Let us recall some essential facts. Some of the basic processes in one loop $SU(3)$ ChPT that were
first studied were form factors that enter into weak decays of mesons. These are readily produced by AMPC
by providing as input the kaon, pion and the $W$-boson, and the kaon, $\eta$ and the $W$-boson.
These when properly normalized yield the $K\pi$ and $K\eta$ form factors.  We have
checked the amplitudes from AMPC and we present the results.

Of the basic meson processes, the earliest to have been computed are the
$K\pi$ \cite{Bernard:1990kw} and $\pi\eta$ scattering amplitudes \cite{Bernard:1991xb}. The
$\pi\pi$\cite{Bernard:1991zc} amplitude was also computed
by these authors.  The two $KK$ amplitudes were computed by Guererro
and Oller \cite{Guerrero:1998ei}.  These have all been
collected by Gomez-Nicola and Pelaez (GNP) \cite{GomezNicola:2001as}.  In addition they
computed the three remaining amplitudes, the $K\eta$ elastic, $K\eta\rightarrow K\pi$, and $\eta\eta$ scattering
amplitudes.  Here we explicitly provide notebooks that produce the results from AMPC.  We have checked all
the amplitudes in GNP and find complete agreement, when the Gell-Mann-Okubo (GMO) relation is used both 
in their results as well as in AMPC result.

Another process we have looked at is the Kaon-Compton process which was studied in
\cite{Guerrero:1997rd}, see also \cite{Fuchs:2000pn}. 
By fixing a factor of 4 in $B(t\nu)$ in Ref.~\cite{Guerrero:1997rd},
we bring AMPC and \cite{Guerrero:1997rd} into
agreement. The loop part agrees and we do not repeat it here. 
Our example is done setting the AMPC switch ``onlytreep2 = 1''.

It is possible to employ AMPC to compute several amplitudes in the
odd-intrinsic parity sector or the anomaly sector.  We have carried out what we believe to be a
comprehensive test of AMPC accessible amplitudes that are available in the literature. Of special interest are
the non-leptonic kaon decays. We verify the results expressed in Table 1 of Ecker et al.\cite{Ecker:1993cq}
and provide the explicit contributions to the amplitudes. In addition, we have generated all the contributions
from the 27-plet contributions.

It should be kept in mind that this report is to serve primarily
as a user manual-cum-report on checks carried out. It is not meant to be
a comprehensive review of existing results. We also provide references to those published
works with which our comparisons have been made, which are not often the first to report results.
Earlier references may be traced from those.

\section{Chiral amplitudes and Form Factors \label{sec:chiral}}

In this section, we present various checks and results obtained with AMPC. 
Here all the external particles, including final states are treated as incoming and
hence the signs of the momenta are labeled accordingly while writing out the momentum
conservation for each process. We give all these specifications along with the
associated scalar products explicitly in the input notebooks.
As mentioned in Sec.\ref{sec:intro}, we caution the reader that
AMPC has originally been written in Mathematica$^{\textrm{\scriptsize\copyright}}$ 5 due to which the
older subroutine gives null result for some processes
owing to possible incompatibility of Mathematica$^{\textrm{\scriptsize\copyright}}$ fonts. 
We deposit a new version\footnote{We thank Gerhard Ecker for providing us both version. 
At present, the new version is posted on his home page.}
of the subroutine which was made available 
later and have been added to the ancillary files of the arXiv submission.
One of these two versions reproduce the result, for instance, in the
case of the odd-parity sector $\pi^+ \rightarrow l^+ \nu\gamma$, only the 
new subroutine reproduces the required result while the old one does not. 
It may thus be noted that in this comprehensive study which have been carried out, 
we have found one or another version that yields the results, 
although a priori we could not say which would work. 
In what follows, unless otherwise mentioned, we use the old subroutine for the various processes
under study. To our knowledge both versions give identical in case where
no Lev-Civita symbol is involved, i.e in pure even-intrinsic parity sector. 
We will indicate explicitly whenever we use the new subroutine.
Further, we compare our AMPC results with those in the literature whenever
available.

\subsection{Odd-intrinsic parity sector \label{sec:anomalous}}

\subsubsection{$\pi^0 \rightarrow \gamma\gamma$}  
The AMPC input for this process is given as 
\beq
\left\{\pi _0\left(p_1\right),\gamma \left(k_1\right),\gamma
\left(k_2\right)\right\} \nonumber
\eeq
The anomalous term contributing to the total amplitude for this process reads 
\beq
-\frac{e^2 \left(k_1\right)_{\xi } \left(p_1\right)_{\rho } \epsilon ^{\xi \rho
   \sigma \tau } \epsilon \left(k_1\right)_{\sigma } \epsilon
   \left(k_2\right)_{\tau }}{4 \pi ^2 F_{\pi }}
\eeq

The AMPC Mathematica$^{\textrm{\scriptsize\copyright}}$ notebooks containing the above expressions are given in {\bf
Ip01.nb} and {\bf Op01.nb}. We have checked our result with the  expression given in Eq.~ (5.1), section VI of Donoghue
et.al \cite{Donoghue:1992dd} and Eq.~(160) of Ref.~\cite{Bijnens:1993xi}. 
We caution the reader there is a missing factor of $i$ in the AMPC result 
compared to that of the established result found in \cite{Donoghue:1992dd}. 

\subsubsection{$\eta_8 \rightarrow \gamma\gamma^*$}
The AMPC input for this process is given as 
\beq
\left\{\eta _8(p_1),\gamma (p_3),\gamma (p_2)\right\} \nonumber
\eeq
The anomalous term contributing to the total amplitude for this process reads 
\beq
-\frac{e^2 (p_1)_{\xi } (p_3)_{\rho } \epsilon ^{\xi \rho \sigma \tau }
   \epsilon (p_2)_{\sigma } \epsilon (p_3)_{\tau }}{4 \sqrt{3} \pi ^2
   F_{\pi }}
\eeq

The AMPC Mathematica$^{\textrm{\scriptsize\copyright}}$ notebooks containing the above expressions are given in {\bf
Ip02.nb} and {\bf Op02.nb}. We have checked our result with the  expression given in Eq.~ (160) of 
ref.\cite{Bijnens:1993xi}. Here also, there is a missing factor of $i$ in the AMPC result 
compared to that of the established result found in \cite{Bijnens:1993xi}.

\subsubsection{$\pi^+ \rightarrow l^+ \nu\gamma$}
The AMPC input for this process is given as 
\beq
\left\{\pi _+(k),W_-(Q),\gamma (r)\right\} \nonumber
\eeq
The anomalous term contributing to the total amplitude for this process reads
\beq
\frac{e G_F k_{\xi } l_{\mu } Q_{\rho } \hat{V}_{\text{ud}} \epsilon ^{\xi \rho \mu
   \tau } \epsilon (r)_{\tau }}{8 \pi ^2 F_{\pi }}
\eeq

The AMPC Mathematica$^{\textrm{\scriptsize\copyright}}$ notebooks containing the above expressions are given in {\bf
Ip03.nb} and {\bf Op03.nb}. We have checked our result against the  expression given in Eq.~ (7) of ref. \cite{Ametller:1993hg}.
We find that our result agrees except for the fact that we need to use the newer version of the AMPC subroutine 
which is given. For comparison purpose, we simplify our results by replacing $Q$ by $-q$ so that the lepton pairs are
outgoing and also making the replacement $f = \sqrt{2}F_\pi$ where $f$ is the pion 
decay constant $f_\pi = f = 132 \,{\rm MeV} (= f_K,$ at lowest order). We have added these remarks to assist the reader
with differing conventions.

\subsubsection{$K^+ \rightarrow l^+ \nu\gamma$}
The AMPC input for this process is given as 
\beq 
\left\{K_+(k),W_-(Q),\gamma (r)\right\} \nonumber
\eeq
The anomalous term contributing to the total amplitude for this process reads
\beq
\frac{e G_F k_{\xi } l_{\mu } Q_{\rho } \hat{V}_{\text{us}} \epsilon ^{\xi \rho \mu
\tau } \epsilon (r)_{\tau }}{8
   \pi ^2 F_{\pi }}
\eeq

The AMPC Mathematica$^{\textrm{\scriptsize\copyright}}$ notebooks containing the above expressions are given in {\bf
Ip04.nb} and {\bf Op04.nb}.
We have checked our result with the  expression given in Eq.~(8) of Ref. \cite{Ametller:1993hg}.
We find that our result agrees except for a factor of $m_K/m_\pi$, which therefore limits the use
of AMPC in this and related process. Here also, we
obtain the 
result only with the newer version of the AMPC subroutine. As in the previous case,
we  
replace $Q$ by $-q$ so that the lepton pairs are outgoing and also make the
replacement $f = \sqrt{2}F_\pi$.

\subsubsection{$\eta_8 \rightarrow \pi^+\pi^-\gamma$}
The AMPC input for this process is given as 
\beq
\left\{\eta _8(p_3),\pi _-(p_1),\pi _+(p_2),\gamma (q)\right\}
\nonumber
\eeq
The anomalous term contributing to the total amplitude for this process reads
\beq
\frac{e (p_1)_{\xi } (p_2)_{\rho } (p_3)_{\sigma } \epsilon ^{\xi \rho
   \sigma \tau } \epsilon (q)_{\tau }}{4 \sqrt{3} \pi ^2 F_{\pi }^3}
\eeq

The AMPC Mathematica$^{\textrm{\scriptsize\copyright}}$ notebooks containing the above expressions are given in {\bf
Ip05.nb} and {\bf Op05.nb}. We have checked our result against the expression given in Eq.~ (2)  of 
Ref.\cite{Ametller:1997qy}. Our results agree.

\subsubsection{$\tau^- \rightarrow \eta_8 \, \pi^-\pi^0\pi^0\nu$}
The AMPC input for this process is given as 
\beq
\left\{W_-(q),\eta _8(k),\pi _+(p_1),\pi _0(p_2),\pi
   _0(p_3)\right\} \nonumber
\eeq
The anomalous term contributing to the total amplitude for this process reads
\beq
\frac{i G_F k_{\xi } l_{\mu } (p_1)_{\rho } q_{\sigma } \hat{V}_{\text{ud}}
   \epsilon ^{\xi \rho \sigma \mu }}{4 \sqrt{3} \pi ^2 F_{\pi }^4}
\eeq

The AMPC Mathematica$^{\textrm{\scriptsize\copyright}}$ notebooks containing the above expressions are given in {\bf
Ip06.nb} and {\bf Op06.nb}.

\subsubsection{$\tau^- \rightarrow \eta_8 \, \pi^-\pi^0\nu$}
The AMPC input for this process is given as 
\beq
\left\{W_-(p),\eta _8(q_1),\pi_+(q_2),\pi_0(q_3)\right\} \nonumber
\eeq
The anomalous term contributing to the total amplitude for this process reads
\beq
-\frac{G_F l_{\mu } p_{\xi }
   (q_1)_{\rho } (q_2)_{\sigma
   } \hat{V}_{\text{ud}} \epsilon ^{\xi
   \rho \sigma \mu }}{4 \sqrt{3} \pi ^2
   F_{\pi }^3}
\eeq

The AMPC Mathematica$^{\textrm{\scriptsize\copyright}}$ notebooks containing the above expressions are given in {\bf
Ip07.nb} and {\bf Op07.nb}. We obtain this result with the newer version of the AMPC subroutine.

\subsubsection{$\tau^- \rightarrow K^- \, \pi^- K^+\nu$}

The AMPC input for this process is given as 
\beq
\left\{W_-(q),K_+\left(k_1\right),\pi _+(p),K_-\left(k_2\right)\right\} \nonumber
\eeq
The anomalous term contributing to the total amplitude for this process reads
\beq
-\frac{G_F \left(k_1\right)_{\sigma } l_{\mu } p_{\xi } q_{\rho } \hat{V}_{\text{ud}}
   \epsilon ^{\xi \rho \sigma \mu }}{4 \pi ^2 F_{\pi }^3}
\eeq

The AMPC Mathematica$^{\textrm{\scriptsize\copyright}}$ notebooks containing the above expressions are given in {\bf
Ip08.nb} and {\bf Op08.nb}. Here also, we obtain the results with the newer version of the AMPC subroutine.

\subsection{$\gamma\pi^- \rightarrow \pi^-\pi^0$}
The AMPC input for this process is given as 
\beq
\left\{\gamma (A),\pi _-\left(p_1\right),\pi _+\left(p_2\right),\pi
   _0\left(p_0\right)\right\} \nonumber
\eeq
The anomalous term contributing to the total amplitude for this process reads
\beq
-\frac{e A_{\xi } \left(p_1\right)_{\rho } \left(p_2\right)_{\sigma } \epsilon
   ^{\xi \rho \sigma \tau } \epsilon (A)_{\tau }}{4 \pi ^2 F_{\pi }^3}
\eeq

The AMPC Mathematica$^{\textrm{\scriptsize\copyright}}$ notebooks containing the above expressions are given in {\bf
Ip09.nb} and {\bf Op09.nb}. We have checked our result against the expression given in  Eq.~(202) of \cite{Bijnens:1993xi}.
We caution the reader there is a missing factor of $i$ in the AMPC result 
compared to that of the established result found in \cite{Bijnens:1993xi}.

\subsection{$K_{l4}$ decay.}
We investigated the anomalous part of the $K_{l4}$
decay $K^+(p)\to l(q_l)\nu(q_\nu)\pi^-(q_2)\pi^+(q_1)$
in AMPC. The AMPC input for the process is

\beq
\left\{K_+(p),W_-(q),\pi _+\left(q_2\right),\pi _-\left(q_1\right)\right\} \nonumber
\eeq

where, $q=q_l+q_\nu$. The anomalous part obtained from AMPC
is 

\beq
-\frac{G_F \hat{V}_{\text{us}}\epsilon^{\xi\rho\sigma\mu}p_\xi q_\rho q_{2\sigma}l_\mu}
{4\pi^2 F_\pi^3}.
\eeq
where $l_\mu$ is the leptonic part of
the amplitude. This result can be compared with Eq.(8) of 
~\cite{Ametller:1993hg}.

It may be noted that the old version of the AMPC does not give the 
anomalous part, and it can be only obtained in the new version of the 
AMPC. The input and output notebook for this process are \textbf{Ip10.nb}
and \textbf{Op10.nb} respectively. This notebook also produces the even-intrinsic
part which agrees with ~\cite{Ametller:1993hg}. Note here that the pole part is correctly
reproduced by AMPC.

\subsection{Chiral anomaly in nonleptonic radiative kaon decays.}
The chiral anomaly in the non-leptonic radiative kaon decays is discussed in detail in
~\cite{Ecker:1993cq}. Such decays can be described by the $\Delta S=1$ 
weak Hamiltonian

\be\label{lagS1}
\mathcal{H}^{\Delta S=1}=\frac{G_F}{\sqrt{2}}V_{ud}V_{us}^*\Sigma_{i}C_i Q_i+hc,~
\ee
where $Q_i$ are the four quark operators and $C_i$ Wilson coefficients.
The Lagrangian~(\ref{lagS1}) has two parts, one that transforms as an octet and
another as a 27-plet under the chiral transformation. The corresponding
coupling constants are $G_8$ and $G_{27}$. The chiral anomaly contributes to
the coefficients $N_{28}, N_{29}, N_{30}, N_{31}$ of the octet operator and to 
the coefficients $R_{21}, R_{22}, R_{23}$ of the 27-plet operator. 
The anomaly contribution coming from the octet and 27-plet part of the amplitude are 
isolated by separating the coefficients of $N$ and $R$ respectively.
It may be noted that in AMPC the coupling constant for the octet and the 27-plet part for $K^0$ decay 
are $G_8$ and $G_{27}$, and for a $\overline{K}^0$ decay they are called $\hat G_8$ and 
$\hat G_{27}$. However in the limit of $CP$ conservation $\hat G_8\to G_8$ and $\hat G_{27}\to G_{27}$.
Using AMPC, we calculate the amplitude of $K^{+0}$ decay into two and three pions and
a photon and the anomaly part of the amplitude is checked against the 
Table 1 of~\cite{Ecker:1993cq}. It may be noted that the explicit amplitudes
of all the AMPC accessible decays considered here are not given in~\cite{Ecker:1993cq}.
However we find agreement with \cite{Ecker:1993cq} regarding the anomaly contributions 
coming from different octet and 27-plet operators.

\subsubsection{Anomaly contribution to $K^+\to \pi^+\pi^0\gamma$ decay.}
We consider the process $K^+(k)\to \pi^+(p_1)\pi^0(p_2)\gamma(q)$
The anomaly contribution coming from the octet part of the contribution is

\be
\mathcal{A}^8 =\frac{8 e G_8 k_{\xi } (N_{30}^r-3 N_{29}^r) 
p_{1\rho } p_{2\sigma } \epsilon ^{\xi \rho \sigma \tau } \epsilon(q)_\tau}
{F_\pi}
\ee

The anomaly coming from the 27-plet part of the amplitude is
given by

\be
\mathcal{A}^{27} =\frac{2 e G_{27} k_{\xi } p_{1\rho } p_{2\sigma } \left(5 R_{22}^r-3 R_{23}^r\right) \epsilon ^{\xi \rho \sigma \tau } \epsilon
   (q)_{\tau }}{3 F_{\pi }}
\ee

The input and output notebooks for this process are given in \textbf{Ip11.nb}
\textbf{Op11.nb} respectively.

\subsubsection{Anomaly contribution to $K^+\to \pi^+\pi^0\pi^0\gamma$ decay.}
The anomaly contribution coming from the octet part of the Lagrangian
for the process $K^+(p)\to \pi^+(q)\pi^0(r)\pi^0(s)\gamma(t)$

\be
\mathcal{A}^8 =\frac{4 i e G_8 \left(N_{30}^r-3 N_{29}^r\right) p_{\xi } q_{\rho } 
\epsilon ^{\xi \rho \sigma \tau } \left(r_{\sigma }+s_{\sigma }\right) \epsilon
   (t)_{\tau }}{F_{\pi }^2}
\ee
and the contribution coming from the 27-plet part is
\be
\mathcal{A}^{27} =\frac{i e G_{27} p_{\xi } q_{\rho } 
\left(5 R_{22}^r-3 R_{23}^r\right) \epsilon ^{\xi \rho \sigma \tau } 
\left(r_{\sigma }+s_{\sigma }\right) \epsilon
   (t)_{\tau }}{3 F_{\pi }^2}
\ee

The input and the output notebooks for this process are given in \textbf{Ip12.nb}
\textbf{Op12.nb} respectively.

\subsubsection{Anomaly contribution to $K^+\to \pi^+\pi^+\pi^-\gamma$ decay.}
We considered the process $K^+(p_1)\to \pi^+(p_2)\pi^+(p_3)\pi^-(p_4)\gamma(q)$.
The anomaly contribution coming from the octet part of the Lagrangian
is

\be
\mathcal{A}^8=\frac{16 i e G_8 \left(N_{29}^r+N_{31}^r\right) p_{1\xi } 
\left(p_{2\rho }+p_{3\rho }\right) p_{4\sigma } \epsilon ^{\xi \rho \sigma \tau } \epsilon
   (q)_{\tau }}{F_{\pi }^2}
\ee

The anomaly contribution from the 27-plet part is

\be
\mathcal{A}^{27}=\frac{8 i e G_{27} p_{1\xi } 
\left(p_{2\rho }+p_{3\rho }\right) p_{4\sigma } 
\left(R_{22}^r-3 R_{23}^r\right) \epsilon ^{\xi \rho \sigma \tau }
   \epsilon (q)_{\tau }}{3 F_{\pi }^2}
\ee

The input and the output notebooks for this process are given in \textbf{Ip13.nb}
\textbf{Op13.nb} respectively.

\subsubsection{Anomaly contribution to $K_{L,S}\to \pi^+\pi^-\pi^0\gamma$ decay.}
In the limit of $CP$ conservation we can write the $K_L$ and $K_S$
as

\begin{eqnarray}
 K_L &=&\frac{1}{\sqrt{2}}(K^0+\overline{K}^0)\nn\\
 K_S &=&\frac{1}{\sqrt{2}}(K^0-\overline{K}^0)\nn
\end{eqnarray}

Using AMPC the decay $K_0(k),\overline{K}_0(k)\to \pi^+(-p)\pi^-(-p_2)\pi^0(-p_3)\gamma(-q)$,
are calculated and the anomaly part of the amplitude is separated.
By adding and subtracting, we get respectively the anomaly
contributions of the amplitude in $K_L$ and $K_S$ decays.

\bea
\mathcal{A}_{K_L}^{G_8}&=&\frac{4 i \sqrt{2} e G_8 k_{\xi } \left(6 N_{28}^r+3 N_{29}^r-5 N_{30}^r\right) \left(p_{1\rho
   }+p_{2\rho }\right) p_{3\sigma } \epsilon ^{\xi \rho \sigma \tau } \epsilon (q)_{\tau }}{F_{\pi
   }^2}\nn\\
\mathcal{A}_{K_S}^{G_8}&=&\frac{4 i \sqrt{2} e G_8 N_{29}^r \epsilon ^{\xi \rho \sigma \tau } \epsilon (q)_{\tau } (p_{3\sigma }
   \left(p_{2\rho } \left(5 k_{\xi }+8 p_{1\xi }\right)-5 k_{\xi } p_{1\rho }\right)-2
   k_{\xi } p_{1\rho } p_{2\sigma })}{F_{\pi }^2}\nn\\&+&\frac{4 i \sqrt{2} e G_8 N_{30}^r
   \left(p_3\right)_{\sigma } \epsilon ^{\xi \rho \sigma \tau } \epsilon (q)_{\tau } \left(k_{\xi } \left(p_1\right)_{\rho
   }-\left(p_2\right)_{\rho } \left(k_{\xi }+2 \left(p_1\right)_{\xi }\right)\right)}{F_{\pi }^2}\nn\\&-&\frac{8 i \sqrt{2} e G_8
   N_{31}^r \epsilon ^{\xi \rho \sigma \tau } \epsilon (q)_{\tau } \left(k_{\xi } \left(p_1\right)_{\rho }
   \left(p_2\right)_{\sigma }+\left(p_3\right)_{\sigma } \left(k_{\xi } \left(p_1\right)_{\rho }-\left(p_2\right)_{\rho }
   \left(k_{\xi }+\left(p_1\right)_{\xi }\right)\right)\right)}{F_{\pi }^2}\nn
\eea

\bea
\mathcal{A}_{K_L}^{G_{27}}&=&\frac{4 i \sqrt{2} e G_{27} k_{\xi } \left(\left(p_1\right)_{\rho }+\left(p_2\right)_{\rho }\right) \left(p_3\right)_{\sigma }
   R_{21}^r \epsilon ^{\xi \rho \sigma \tau } \epsilon (q)_{\tau }}{3 F_{\pi }^2}\nn\\&+&\frac{i \sqrt{2} e G_{27} k_{\xi } R_{23}^r
   \epsilon ^{\xi \rho \sigma \tau } \left(\left(p_1\right)_{\rho } \left(2 \left(p_2\right)_{\sigma }-7 \left(p_3\right)_{\sigma
   }\right)-9 \left(p_2\right)_{\rho } \left(p_3\right)_{\sigma }\right) \epsilon (q)_{\tau }}{3 F_{\pi }^2}\nn\\&-&\frac{i \sqrt{2} e
   G_{27} k_{\xi } R_{22}^r \epsilon ^{\xi \rho \sigma \tau } \left(25 \left(p_2\right)_{\rho } \left(p_3\right)_{\sigma
   }+\left(p_1\right)_{\rho } \left(6 \left(p_2\right)_{\sigma }+31 \left(p_3\right)_{\sigma }\right)\right) \epsilon (q)_{\tau
   }}{3 F_{\pi }^2}\nn\\
\mathcal{A}_{K_S}^{G_{27}}&=&-\frac{4 i \sqrt{2} e G_{27} k_{\xi } \left(\left(p_1\right)_{\rho }+\left(p_2\right)_{\rho }\right) \left(p_3\right)_{\sigma }
   R_{20}^r \epsilon ^{\xi \rho \sigma \tau } \epsilon (q)_{\tau }}{3 F_{\pi }^2}\nn\\&+&\frac{i \sqrt{2} e G_{27}
   \left(p_3\right)_{\sigma } R_{23}^r \epsilon ^{\xi \rho \sigma \tau } \epsilon (q)_{\tau } \left(5 k_{\xi }
   \left(p_1\right)_{\rho }+\left(p_2\right)_{\rho } \left(9 k_{\xi }+4 \left(p_1\right)_{\xi }\right)\right)}{3 F_{\pi
   }^2}\nn\\&-&\frac{i \sqrt{2} e G_{27} R_{22}^r \epsilon ^{\xi \rho \sigma \tau } \epsilon (q)_{\tau } \left(8 k_{\xi }
   \left(p_1\right)_{\rho } \left(p_2\right)_{\sigma }+\left(p_3\right)_{\sigma } \left(19 k_{\xi } \left(p_1\right)_{\rho
   }+\left(p_2\right)_{\rho } \left(7 k_{\xi }-4 \left(p_1\right)_{\xi }\right)\right)\right)}{3 F_{\pi }^2}\nn
\eea

The input and outpus notebooks for the processes $K_0\to \pi^+\pi^-\pi^0\gamma$ 
and $\overline{K}_0\to \pi^+\pi^-\pi^0\gamma$ can be found
in  \textbf{Op14.nb} and \textbf{Op14.nb} respectively. We have 
extracted the anomalous parts of the amplitude 
$K_{L,S}\to \pi^+\pi^-\pi^0\gamma$ in the  same notebook.

\subsubsection{Anomaly contribution in $K_{L,S}\to \gamma\pi^-\pi^+$}
We calculate the anomaly contribution in $K_{L,S}\to \gamma\pi^-\pi^+$
in the same way as we did in the previous sections.
Using AMPC we calculate the decay $K_0(p)\to \gamma(-q)\pi^-(-p_1)\pi^+(-p_2)$,
and $\overline{K}_0(p)\to \gamma(-q)\pi^-(-p_1)\pi^+(-p_2)$
and extract the anomaly parts coming from the octet and 27-plet part
of the Lagrangian. We finally add and subtract these anomaly parts
to obtain the anomaly contribution to $K_{L,S}\to \gamma\pi^-\pi^+$ decay.

\bea
\mathcal{A}_{K_L}^{G_8}&=&\frac{16 \sqrt{2} e G_8 N_{29}^r p_{\xi } \left(p_1\right)_{\sigma } q_{\rho } \epsilon ^{\xi \rho \sigma \tau } \epsilon
   (q)_{\tau }}{F_{\pi }}+\frac{16 \sqrt{2} e G_8 N_{31}^r p_{\xi } \left(p_1\right)_{\sigma } q_{\rho } \epsilon ^{\xi \rho
   \sigma \tau } \epsilon (q)_{\tau }}{F_{\pi }}\nn\\
\mathcal{A}_{K_S}^{G_8}&=&0\nn
\eea

\bea
\mathcal{A}_{K_L}^{G_{27}}&=&\frac{16 \sqrt{2} e G_{27} p_{\xi } \left(p_1\right)_{\sigma } q_{\rho } R_{22}^r \epsilon ^{\xi \rho \sigma \tau } \epsilon
   (q)_{\tau }}{3 F_{\pi }}\nn\\
\mathcal{A}_{K_S}^{G_{27}}&=&\frac{4 \sqrt{2} e G_{27} p_{\xi } \left(p_1\right)_{\sigma } q_{\rho } R_{22}^r \epsilon ^{\xi \rho \sigma \tau } \epsilon
   (q)_{\tau }}{F_{\pi }}-\frac{4 \sqrt{2} e G_{27} p_{\xi } \left(p_1\right)_{\sigma } q_{\rho } R_{23}^r \epsilon ^{\xi \rho
   \sigma \tau } \epsilon (q)_{\tau }}{3 F_{\pi }}\nn
\eea

The input and output notebooks for the processes $K_0\to \gamma\pi^-\pi^+$ 
and $\overline{K}_0\to \gamma\pi^-\pi^+$ can be found
in  \textbf{Ip15.nb} and \textbf{Op15.nb} respectively. We have 
extracted the anomalous parts of the amplitude 
$K_{L,S}\to \gamma\pi^-\pi^+$ in the  same notebook.

\subsection{Form Factors results from AMPC}
We present a check for the $\pi^+\pi^+$, $K^+K^-$, $K^0K^0$, $K^+\eta$ and the
$K^+\pi^0$ form factors given in Eq.~(2.1) of Gasser et.al., \cite{Gasser:1984ux}. The matrix elements are defined below
\bea
\langle \pi^+|j_\mu|\pi^+\rangle &=& (p_\mu^{'}+p_\mu)F_V^\pi(t), \nonumber \\
\langle K^+|j_\mu|K^+\rangle &=& (p_\mu^{'}+p_\mu)F_V^{K^+}(t), \nonumber \\
\langle K^0|j_\mu|K^0\rangle &=& (p_\mu^{'}+p_\mu)F_V^{K^0}(t), \nonumber \\
\langle K^+|\overline{u}\gamma_\mu s|\eta \rangle &=&
\sqrt{\frac{3}{2}}[(p_\mu^{'}+p_\mu)f_+^{K\eta}(t) +
(p_\mu^{'}-p_\mu)f_-^{K\eta}(t)], \nonumber \\
\langle K^+|\overline{u}\gamma_\mu s|\pi^0 \rangle &=&
\sqrt{\frac{1}{2}}[(p_\mu^{'}+p_\mu)f_+^{K\pi}(t)+ (p_\mu^{'}-p_\mu)f_-^{K\pi}(t)]
\nonumber \\
\eea

The above processes are AMPC accessible since they appear in semi-leptonic weak decays.
The form factors appear in the above matrix elements which are denoted by $f_+^{K\pi}(t)$,
$f_-^{K\pi}(t)$,$f_0^{K\pi}(t)$ and $f_+^{K\eta}(t)$,$f_-^{K\eta}(t)$, $f_0^{K\eta}(t)$ respectively for the
$K^+\pi^0$ and the $K^+\eta$. We have extracted the coefficients for these vector currents from our AMPC amplitudes
for the given processes with appropriate matching of various functions appearing
in both the AMPC and the the results in \cite{Gasser:1984ux}.
We check explicitly against the various form factor definitions given in Eq.(2.4)
of \cite{Gasser:1984ux}.
The AMPC results are given in terms of the $\bar{A}$ and $\bar{B}$ functions where
\bea
\bar{A}[M^2] = -M^2/(4 \pi)^2 {\rm ln}[M^2/\mu^2], \quad \quad  
\bar{B}[t,M_P^2,M_Q^2] = \bar{J}[t,M_P^2,M_Q^2] 
\eea
with the $\bar{J}[t,M_P^2,M_Q^2]$ as given in \cite{Gasser:1984gg}.
We present the various definitions found in \cite{Gasser:1984gg, Gasser:1984ux}
required for 
evaluating the form factors.
\bea
H_{PQ}(t) &=& \frac{1}{F_0^2}(tM^r(t)-L(t))+\frac{2}{3F_0^2}L_9^rt, \\
M^r(t) &=&
\frac{1}{12t}(t-2\Sigma)\bar{J}(t)+\frac{\Delta^2}{3t^2}\bar{\bar{J}}(t)-\frac{1}{6}k+\frac{1}{288\pi^2},
\eea
where,
\begin{equation}
 \Sigma = M_P^2 + M_Q^2, \quad \quad \quad \quad \Delta = M_P^2 - M_Q^2, \quad \quad
\quad \quad L(t) = \frac{\Delta^2}{4t}\bar{J}
\end{equation}

Two cases arise where the form factors could be for the equal mass like in the case
of the $\pi\pi$ and $KK$ while they could be for the unequal mass case like in the
$K^+\pi^0$ and the $K^+\eta$ \footnote{The standard loop functions can be found in
\cite{Gasser:1984gg, Gasser:1984ux}. They are given here in the interest of making this paper fully self-contained.
The reader may always consult primary references for further clarification if required.}.

\bigskip

{\bf Case 1:} $(M_P = M_Q = M)$,
\beq
 \bar{J}(t) = \frac{1}{16\pi^2}\left(\sigma(t){\rm ln}\frac{\sigma(t)-1}{\sigma(t)+1}+2\right),
\eeq
where,
\begin{equation}
 \sigma(t) = \sqrt{1-\frac{4M^2}{t}},
\end{equation}
Also, 
\begin{equation}
 \bar{\bar{J}}(t) = \bar{J}(t)-t\bar{J'}(0),
\end{equation}
\begin{equation}
 \bar{J'}(0) = \frac{1}{96\pi^2}\frac{1}{M^2},
\end{equation}
\begin{equation}
 k = \frac{1}{32\pi^2}\left({\rm ln}\frac{M^2}{\mu^2}+1\right)
\end{equation}

\bigskip

{\bf Case 2:} $(M_P \neq M_Q)$,
\beq
\bar{J}(t) = \frac{1}{32\pi^2}\left(2+\frac{\Delta}{t}{\rm ln}\frac{M_Q^2}{M_P^2} - \
\frac{\Sigma}{\Delta}{\rm ln}\frac{M_Q^2}{M_P^2} - \frac{\nu}{t}{\rm
ln}\frac{(t+\nu)^2-\Delta^2}{t-\nu)^2-\Delta^2}\right)
\eeq
where,
\begin{equation}
\nu^2 = t-(M_P+M_Q)^2)(t-(M_P-M_Q)^2)
\end{equation}
Also,
\begin{equation}
 \bar{J'}(0) = \frac{1}{32\pi^2}\left(\frac{\Sigma}{\Delta^2} +
2\frac{M_P^2M_Q^2}{\Delta^3}{\rm ln}\frac{M_Q^2}{M_P^2}\right)
\end{equation}

\begin{equation}
 k = \frac{1}{32\pi^2}\left(M_P^2{\rm ln}\frac{M_P^2}{\mu^2}-M_Q^2{\rm
ln}\frac{M_Q^2}{\mu^2}\right)\frac{1}{\Delta}
\end{equation}

\bigskip

We recall that the matrix element for the weak decay is given by
\begin{equation}\label{matrix}
 \mathcal{M} =-\frac{G_F}{\sqrt{2}}V_\text{CKM}j_\mu J^\mu 
\end{equation}

where, $j_\mu$ and $J^\mu$ are the leptonic and hadronic currents
respectively, and $V_{\text{CKM}}$ is the CKM elements. In AMPC the weak decay amplitude 
is written with a vertex factor $-G_F$. To match the AMPC conventions 
with that of in literatures, we have multiplied
the AMPC results with $\sqrt{2}$.

The input and output notebooks for the form factors of different processes considered
are tabulated in Table ~\ref{tableA}.

\vspace{0.4cm}

\begin{table*}[!ht]
\begin{center}
\caption{Input and output notebooks for the form factors.}
\vspace{0.5cm}
\label{tableA}
 \begin{tabular}{ l || c | c  }
\hline 
Process & Input Notebook & Output Notebook \\ [1ex]
\hline
$\pi^+\pi^+$ & \textbf{Ip16a.nb} & \textbf{Op16a.nb} \\ [1ex]
\hline
$K^+K^-$ & \textbf{Ip16b.nb} & \textbf{Op16b.nb} \\ [1ex]
\hline
$K^0K^0$ & \textbf{Ip16c.nb} & \textbf{Op16c.nb} \\ [1ex]
\hline
$k^+\eta$ & \textbf{Ip16d.nb} & \textbf{Op16d.nb} \\ [1ex]
\hline
$K^+\pi^0$ & \textbf{Ip16e.nb} & \textbf{Op16e.nb} \\ [1ex]
\hline
\end{tabular}
\end{center}
\end{table*}

Using the above expressions for the equal mass case
and doing the necessary simplifications, we find that our AMPC result 
agrees for the equal mass case. The comparison and simplification is
given in detail in the AMPC notebook {\bf Op16equal.nb}.
Coming to the unequal mass case, the $f_+(t)$ agrees for both the $K\pi$ and
the $K\eta$ for the expression given in \cite{Gasser:1984ux}. The comparison is
given in detail in the AMPC notebook {\bf Op16unequal.nb, Op16d1.nb, Op16e1.nb}.
We check $f_-(t)$ for the $K\pi$ form factor using the expression given
in in Eq.~ (4.4) of \cite{Bijnens:2003uy} against our result and find that they agree.
As a check we also do the calculation for $f_+(t)$ given in Eq.(4.3) and find
agreement. For details of comparison, see the AMPC notebook {\bf Op16e2.nb}.
In doing this calculation, the expressions for the various loop functions are taken
from 
\cite{Unterdorfer:2005au} which are introduced below.
\bea
\bar{B}_{20}(t) &=& -\frac{t-3M_P^2-3M_Q^2}{288\pi^2} + \frac{\bar{A}(M_Q^2) +
2\bar{B}(t)M_P^2 - (M_P^2-M_Q^2+t)\bar{B}_{11}(t)}{6} \nonumber \\
\bar{B}_{22}(t) &=& \frac{t-3M_P^2-3M_Q^2}{288\pi^2t} + \frac{\bar{A}(M_Q^2) -
\bar{B}(t)M_P^2 + 2(M_P^2-M_Q^2+t)\bar{B}_{11}(t)}{3t} \nonumber\\
\bar{B}_{11}(t) &=& \frac{-\bar{A}(M_P^2) + \bar{A}(M_Q^2) +
\bar{B}(t)(M_P^2-M_Q^2+t)}{2t}\nonumber
\eea

\beq
{B}(t) = \bar{B}(t) + B(0)
\eeq
where,
\bea 
\bar{B}(t) = \frac{1}{32\pi^2}(2 + \frac{M_P^2-M_Q^2}{t}{\rm ln}\frac{M_Q^2}{M_P^2}
- \frac{M_P^2+M_Q^2}{M_P^2-M_Q^2}{\rm ln}\frac{M_Q^2}{M_P^2} - 
\frac{\sqrt{\lambda(t,M_P^2,M_Q^2)}}{t} \nonumber\\
{\rm ln}\frac{(t+\sqrt{\lambda(t,M_P^2,M_Q^2)})^2 -
(M_P^2-M_Q^2)^2}{(t-\sqrt{\lambda(t,M_P^2,M_Q^2)})^2 - (M_P^2-M_Q^2)^2})
\eea
\beq
B(0) = \frac{\bar{A}(M_P^2) - \bar{A}(M_Q^2)}{M_P^2-M_Q^2}
\eeq

It may be noted that the notations in \cite{Bijnens:2003uy} are different from that
of \cite{Unterdorfer:2005au}.  Specifically we give the relations,
$\bar{B}_{22}$\cite{Bijnens:2003uy} = $\bar{B}_{20}$\cite{Unterdorfer:2005au}, 
$\bar{B}_{21}$\cite{Bijnens:2003uy} = $\bar{B}_{22}$\cite{Unterdorfer:2005au}, 
$\bar{B}$\cite{Bijnens:2003uy} = $B$\cite{Unterdorfer:2005au}. This is done by
comparing the Lorentz structures in the expressions (B.1) of \cite{Bijnens:2002hp}
with eqn (B.5) of
\cite{Unterdorfer:2005au}.

\subsection{Amplitude for Kaon polarizability $\gamma K^+ \rightarrow \gamma K^+$}
One of the AMPC applications is of special interest to studying the Compton
amplitudes.  The pion amplitude
was computed by Bijnens et.al.,\cite{Bijnens:1987dc},
while the kaon analog was computed by Guererro and Prades \cite{Guerrero:1997rd} and
later by
Fuchs et.al.,\cite{Fuchs:2000pn}. The amplitude for the process  
$\gamma(q_1) K^+(p_1) \rightarrow \gamma(q_2) K^+(p_2)$ is given in
terms of $A(t,\nu)$ and $B(t,\nu)$. The tree level expressions of
$A(t,\nu)$ and $B(t,\nu)$ at $\mathcal{O}(p^4)$ are given  

\begin{eqnarray}
 A(t,\nu) &=& \frac{2}{t-\nu}+\frac{2}{t+\nu}\nn\\
 B(t,\nu) &=& \frac{1}{t}\Big (\frac{1}{t-\nu}+\frac{1}{t+\nu}\Big ),\nn
\end{eqnarray}
where $t$ and $\nu$ are kinematics variables defined in Ref.~\cite{Guerrero:1997rd}.

The amplitude can be generated with the attached input file \textbf{Ip17.nb}.
The output can be found in the attached output file \textbf{Op17.nb},
where we have shown that the tree level amplitude of ~\cite{Guerrero:1997rd} 
does not match with that generated by AMPC, unless the $B(t,\nu)$ is multiplied
by factor 4. The correct expression is given below -
\begin{equation}
 B(t,\nu) = \frac{4}{t}\Big (\frac{1}{t-\nu}+\frac{1}{t+\nu}\Big ),\nn
\end{equation}

It may be noted that in the attached input and output notebooks we have
considered the process $\gamma(k_1) K^+(p_1) \rightarrow \gamma(k_2) K^+(p_2)$.

We have also compared the tree level amplitude of $\gamma K \rightarrow \gamma K$
generated from AMPC against that given in Ref.~\cite{Fuchs:2000pn} and the results agree
providing a further check to the new expression for $B(t,\nu)$ given above.

\subsection{Scattering amplitudes at 1-loop}

As mentioned in Sec. \ref{sec:intro}, we have checked all the processes given in 
GMO \cite{GomezNicola:2001as}, against our AMPC 
results  and we find that they agree to the best of our knowledge. All the notation in
\cite{GomezNicola:2001as} except for the $\mu_\pi, \mu_K, \mu_\eta$ function agrees with the ones
present in AMPC results which are the expressions already introduced.
One crucial simplification needs to be done for the expression of $\overline{J}'(0)$.
This is as follows - 
\bea
\overline{J}'(0) &=& \frac{1}{32\pi^2}[\frac{\Sigma}{\Delta^2} +
3\frac{M_P^2M_Q^2}{\Delta^3}{\rm log}\frac{M_Q^2}{M_P^2}] \\
                 &=& \frac{1}{32\pi^2}\frac{\Sigma}{\Delta^2} +
\frac{M_P^2}{\Delta^3}\overline{A}[M_Q^2] +
\frac{M_Q^2}{\Delta^3}\overline{A}[M_P^2] \\
\overline{A}[M_i^2] &=& - 2 F_\pi^2 \,\, \mu_i \quad \quad \quad \quad i = \pi,K,\eta\\
\mu_i &=& \frac{M_i^2}{32\pi^2F_\pi^2}{\rm log}\frac{M_i^2}{\mu^2} \quad \quad \quad \quad i = \pi,K,\eta
\eea 

As an example we demonstrate our comparison for one of the processes. See attached notebook
{\bf Op18ccheck.nb}. In the results to follow, we use the GMO mass formula,
\be
3M_\eta^2=4M_K^2-M_\pi^2\nn
\ee
as well as appropriate s,t,u relations wherever necessary.

The input and output notebooks for different scattering processes 
are tabulated in Table.~\ref{table}

\begin{table*}[!ht]
\begin{center}
\caption{Input and output notebooks 
for various scattering processes.}
\vspace{0.5cm}
\label{table}
 \begin{tabular}{ l || c | c  }
\hline 
Process & Input Notebook & Output Notebook \\ [1ex]
\hline
$\eta\eta\to\eta\eta$ & \textbf{Ip18a.nb} & \textbf{Op18a.nb} \\ [1ex]
\hline
$\overline{K}^0\eta\to\overline{K}^0\eta$ & \textbf{Ip18b.nb} & \textbf{Op18b.nb} \\ [1ex]
\hline
$\overline{K}^0\eta\to\overline{K}^0\pi^0$ & \textbf{Ip18c.nb} & \textbf{Op18c.nb} \\ [1ex]
\hline
$\overline{K}^0K^0\to K^+ K^-$ & \textbf{Ip18d.nb} & \textbf{Op18d.nb} \\ [1ex]
\hline
$K^+\pi^+\to K^+\pi^+$ & \textbf{Ip18e.nb} & \textbf{Op18e.nb} \\ [1ex]
\hline
$\pi^0\eta\to \pi^0\eta$ & \textbf{Ip18f.nb} & \textbf{Op18f.nb} \\ [1ex]
\hline
\end{tabular}
\end{center}
\end{table*}

\subsection{Application of Chiral Dynamics in $\tau$ decays.}
In Ref.~\cite{Aubrecht:1981cr} the tree level amplitudes for $\tau$
decays to multi-meson states are obtained using $SU(3)
\times SU(3)$ Lagrangian in the limit of vanishing quark 
mass for one, two and three meson final states involving
$\pi$, $K$ and $\eta$. 
In this section we compare the AMPC generated 
amplitudes with that given in ~\cite{Aubrecht:1981cr}.
A few points are in order, regarding the comparisons.
We define the weak matrix element which is given in Eq.~(\ref{matrix}). 
Here the results are presented
up to an overall factor of $G_{F}V_{\text{CKM}}$.

For hadronic matrix elements involving three final
states hadrons, we have simplified the Lorentz
structures and compared the coefficients of the 
momentum vectors. In ~\cite{Aubrecht:1981cr}
the authors neglect the quark masses in the Lagrangian.
However the meson masses are retained in the propagator. The 
denominator of the AMPC results match
with that of ~\cite{Aubrecht:1981cr}, and the numerator
match when the meson masses are neglected. We have used the 
GMO relation to simplify the numerator in few
cases. Let us again emphasize that the results
presented here supersede that of Ref.~\cite{Aubrecht:1981cr}
when quark masses are no longer neglected.

The two and three meson final states are accessible
in AMPC. In the attached notebook \textbf{Op19.nb}, the AMPC generated output for 
each of the processes are shown and simplifications are
done using the FeynCalc~\cite{Mertig:1990an}. 
Also provided are two three input notebooks
\textbf{Ip19a.nb} ($J_\mu(\pi^+\pi^0)$), 
\textbf{Ip19b.nb} ($J_\mu(\pi^+K^+K^-)$), 
\textbf{Ip19c.nb}($J_\mu(\eta_1\eta_2 K^+)$). 

\newpage

\subsubsection{Hadron Current matrix elements in two mesons final state}

\begin{table*}[!ht]
\begin{center}
\caption{Comparison of hadron matrix element from Ref.~\cite{Aubrecht:1981cr} and
that obtained from AMPC for two mesons in the final state. }
 \begin{tabular}{ l || c | c  }
\hline 
Process & \cite{Aubrecht:1981cr} & AMPC \\ [1ex]
\hline
$J_\mu(\pi^+\pi^0)$ & $\sqrt{2}(p_+-p)_\mu$ & $\sqrt{2}(p_+-p)_\mu$ \\ [1ex]
\hline
$J_\mu(K^+\overline{K}^0)$ & $-(k_+-\bar k)_\mu$ & $-(k_+-\bar k)_\mu$ \\ [1ex]
\hline
$J_\mu(\pi^0K^+)$ & $\frac{1}{\sqrt{2}}(k_+-p)_\mu$ & $\frac{1}{\sqrt{2}}(k_+-p)_\mu$ \\ [1ex]
\hline
$J_\mu(\pi^+K^0)$ & $(k-p_+)_\mu$ & $(k-p_+)_\mu$ \\ [1ex]
\hline
$J_\mu(K^+\eta_8)$ & $\frac{\sqrt{3}}{\sqrt{2}}(k_+-\eta)_\mu$ & $\frac{\sqrt{3}}{\sqrt{2}}(k_+-\eta)_\mu$ \\ [1ex]
\hline
\end{tabular}
\end{center}
\end{table*}

\subsubsection{$J_\mu(\pi^+(p_1) \pi^+(p_2) \pi^-(p_-))$}

\begin{table*}[!ht]
\begin{center}
\caption{Comparison of the coefficients of external hadron momentum
from Ref.~\cite{Aubrecht:1981cr} and from AMPC for the hadron
current $J_\mu(\pi^+(p_1) \pi^+(p_2) \pi^-(p_-))$}
 \begin{tabular}{ l || c | c }
\hline     
  Coefficients & Ref.~\cite{Aubrecht:1981cr} &AMPC\\ 
\hline                
  $p_-^\mu$ 
& 
$\frac{\sqrt{2} \left(-4 M_{\pi }^2-3 p_-\cdot p_1-3 p_-\cdot p_2-6 p_1\cdot p_2\right)}{3 F_{\pi } \left(M_{\pi }^2+p_-\cdot
   p_1+p_-\cdot p_2+p_1\cdot p_2\right)}$ 
& 
$-\frac{\sqrt{2}(M_{\pi }^2+p_-.p_1+p_-.p_2+2 p_1.p_2)}{F_{\pi } \left(M_{\pi }^2+p_-.p_1+p_-.p_2+p_1.p_2\right)}$ \\
  $p_1^\mu$ 
& 
$\frac{\sqrt{2} \left(2 M_{\pi }^2+3 p_-\cdot p_1+3 p_-\cdot p_2\right)}{3 F_{\pi } \left(M_{\pi
   }^2+p_-\cdot p_1+p_-\cdot p_2+p_1\cdot p_2\right)}$
& 
$-\frac{\sqrt{2}(-M_{\pi
   }^2-p_-.p_1-p_-.p_2)}{F_{\pi } \left(M_{\pi }^2+p_-.p_1+p_-.p_2+p_1.p_2\right)}$ \\
  $p_2^\mu$ 
& 
$\frac{\sqrt{2} \left(2 M_{\pi }^2+3 p_-\cdot p_1+3 p_-\cdot p_2\right)}{3 F_{\pi }
   \left(M_{\pi }^2+p_-\cdot p_1+p_-\cdot p_2+p_1\cdot p_2\right)}$
& 
$-\frac{\sqrt{2}(-M_{\pi }^2-p_-.p_1-p_-.p_2)}{F_{\pi } \left(M_{\pi
   }^2+p_-.p_1+p_-.p_2+p_1.p_2\right)}$ \\
\hline
\end{tabular}
\end{center}
\end{table*}


\subsubsection{$J_\mu(\pi^0(p_1) \pi^0(p_2) \pi^+(p_+))$}

\begin{table*}[!ht]
\begin{center}
\caption{Comparison of the coefficients of external hadron momentum
from Ref.~\cite{Aubrecht:1981cr} and from AMPC for the hadron
current $J_\mu(\pi^0(p_1) \pi^0(p_2) \pi^+(p_+))$}
 \begin{tabular}{ l || c | c }
\hline     
  Coefficients & Ref.~\cite{Aubrecht:1981cr} & AMPC\\ 
\hline                
  $p_1^\mu$ 
& 
$\frac{\sqrt{2} \left(-2 M_{\pi }^2-3 p_+\cdot p_1-3 p_+\cdot p_2\right)}
{3 F_{\pi } \left(M_{\pi }^2+p_+\cdot p_1+p_+\cdot p_2+p_1\cdot
   p_2\right)}$ 
& 
$\frac{-M_{\pi }^2-2 p_+.p_1-2 p_+.p_2}{\sqrt{2} F_{\pi } \left(M_{\pi }^2+p_+.p_1+p_+.p_2+p_1.p_2\right)}$ \\
  $p_2^\mu$ 
& 
$\frac{\sqrt{2} \left(-2 M_{\pi }^2-3 p_+\cdot p_1-3 p_+\cdot p_2\right)}{3 F_{\pi } \left(M_{\pi }^2+p_+\cdot p_1+p_+\cdot
   p_2+p_1\cdot p_2\right)}$
& 
$\frac{-M_{\pi }^2-2 p_+.p_1-2 p_+.p_2}{\sqrt{2}
   F_{\pi } \left(M_{\pi }^2+p_+.p_1+p_+.p_2+p_1.p_2\right)}$ \\
  $p_+^\mu$ 
& 
$\frac{\sqrt{2} \left(4 M_{\pi }^2+3 p_+\cdot p_1+3 p_+\cdot p_2+6 p_1\cdot p_2\right)}{3 F_{\pi } \left(M_{\pi
   }^2+p_+\cdot p_1+p_+\cdot p_2+p_1\cdot p_2\right)}$
& 
$\frac{3 M_{\pi }^2+2 p_+.p_1+2 p_+.p_2+4 p_1.p_2}{\sqrt{2} F_{\pi } \left(M_{\pi}^2+p_+.p_1+p_+.p_2+p_1.p_2\right)}$ \\
\hline
\end{tabular}
\end{center}
\end{table*}

\newpage

\subsubsection{$J_\mu(\pi^+(p_+) K^+(k_+) K^-(k_-))$}

\begin{table*}[!ht]
\begin{center}
\caption{Comparison of the coefficients of external hadron momentum
from Ref.~\cite{Aubrecht:1981cr} and from AMPC for the hadron
current $J_\mu(\pi^+(p_+) K^+(k_+) K^-(k_-))$}
 \begin{tabular}{ l || c | c }
\hline     
  Coefficients & Ref.~\cite{Aubrecht:1981cr} & AMPC\\ 
\hline                
  $p_+^\mu$ 
& 
$\frac{3 k_-\cdot p_++3 k_-\cdot k_++3 M_K^2-M_{\pi }^2}
{3 \sqrt{2} F_{\pi } \left(k_-\cdot p_++k_+\cdot p_++k_-\cdot
   k_++M_K^2\right)}$ 
& 
$-\frac{-k_-.p_+-k_-.k_+-M_K^2}{\sqrt{2} F_{\pi } \left(k_-.p_++k_+.p_++k_-.k_++M_K^2\right)}$ \\
  $k_+^\mu$ 
& 
$\frac{3 k_-\cdot p_++3 k_-\cdot k_++3 M_K^2-M_{\pi }^2}
{3 \sqrt{2} F_{\pi } \left(k_-\cdot p_++k_+\cdot p_++k_-\cdot
   k_++M_K^2\right)}$
& 
$-\frac{-k_-.p_+-k_-.k_+-M_K^2}{\sqrt{2} F_{\pi }
   \left(k_-.p_++k_+.p_++k_-.k_++M_K^2\right)}$ \\
  $k_-^\mu$ 
& 
$\frac{-3 k_-\cdot p_+-6 k_+\cdot p_+-3 k_-\cdot k_+-3 M_K^2-M_{\pi }^2}
{3 \sqrt{2} F_{\pi } \left(k_-\cdot p_++k_+\cdot
   p_++k_-\cdot k_++M_K^2\right)}$
& 
$-\frac{k_-.p_++2 k_+.p_++k_-.k_++M_K^2}{\sqrt{2} F_{\pi }
   \left(k_-.p_++k_+.p_++k_-.k_++M_K^2\right)}$ \\
\hline
\end{tabular}
\end{center}
\end{table*}

\subsubsection{$J_\mu(\pi^+(p_+) K^0(k) \overline{K}^0(\bar k))$}

\begin{table*}[!ht]
\begin{center}
\caption{Comparison of the coefficients of external hadron momentum
from Ref.~\cite{Aubrecht:1981cr} and from AMPC for the hadron
current $J_\mu(\pi^+(p_+) K^0(k) \overline{K}^0(\bar k))$}
 \begin{tabular}{ l || c | c }
\hline     
  Coefficients & Ref.~\cite{Aubrecht:1981cr} & AMPC\\ 
\hline                
  $p_+^\mu$ 
& 
$\frac{3 k\cdot \bar{k}+3 k\cdot p_++3 M_K^2-M_{\pi }^2}
{3 \sqrt{2} F_{\pi } \left(\bar{k}\cdot p_++k\cdot \bar{k}+k\cdot
   p_++M_K^2\right)}$ 
& 
$-\frac{-k.\bar{k}-k.p_+-M_K^2}{\sqrt{2} F_{\pi } \left(\bar{k}.p_++k.\bar{k}+k.p_++M_K^2\right)}$ \\
  $k^\mu$ 
& 
$\frac{-6 \bar{k}\cdot p_+-3 k\cdot \bar{k}-3 k\cdot p_+-3 M_K^2-M_{\pi }^2}
{3 \sqrt{2} F_{\pi } \left(\bar{k}\cdot p_++k\cdot
   \bar{k}+k\cdot p_++M_K^2\right)}$
& 
$-\frac{2 \bar{k}.p_++k.\bar{k}+k.p_++M_K^2}{\sqrt{2}
   F_{\pi } \left(\bar{k}.p_++k.\bar{k}+k.p_++M_K^2\right)}$ \\
  $\bar k^\mu$ 
& 
$\frac{3 k\cdot \bar{k}+3 k\cdot p_++3 M_K^2-M_{\pi }^2}
{3 \sqrt{2} F_{\pi } \left(\bar{k}\cdot p_++k\cdot
   \bar{k}+k\cdot p_++M_K^2\right)}$
& 
$-\frac{-k.\bar{k}-k.p_+-M_K^2}{\sqrt{2} F_{\pi }
   \left(\bar{k}.p_++k.\bar{k}+k.p_++M_K^2\right)}$ \\
\hline
\end{tabular}
\end{center}
\end{table*}

\subsubsection{$J_\mu(\pi^0(p) \overline{K}^0(\bar k) {K}^+(k_+))$}

\begin{table*}[!ht]
\begin{center}
\caption{Comparison of the coefficients of external hadron momentum
from Ref.~\cite{Aubrecht:1981cr} and from AMPC for the hadron
current $J_\mu(\pi^0(p) \overline{K}^0(\bar k) {K}^+(k_+))$}
 \begin{tabular}{ l || c | c }
\hline     
  Coefficients & Ref.~\cite{Aubrecht:1981cr} & AMPC\\ 
\hline                
  $p^\mu$ 
& 
$\frac{p\cdot k_+-p\cdot \bar{k}}{2 F_{\pi } \left(p\cdot \bar{k}+\bar{k}\cdot k_++p\cdot k_++M_K^2\right)}$ 
& 
$\frac{p.k_+-p.\bar{k}}{2 F_{\pi } \left(p.\bar{k}+\bar{k}.k_++p.k_++M_K^2\right)}$ \\
  $\bar k^\mu$ 
& 
$\frac{p\cdot \bar{k}+2
   \bar{k}\cdot k_++3 p\cdot k_++2 M_K^2}{2 F_{\pi } 
\left(p\cdot \bar{k}+\bar{k}\cdot k_++p\cdot k_++M_K^2\right)}$
& 
$\frac{p.\bar{k}+2 \bar{k}.k_++3 p.k_++2 M_K^2}{2
   F_{\pi } \left(p.\bar{k}+\bar{k}.k_++p.k_++M_K^2\right)}$ \\
  $k_+^\mu$ 
& 
$\frac{-3 p\cdot \bar{k}-2
   \bar{k}\cdot k_+-p\cdot k_+-2 M_K^2}{2 F_{\pi } 
\left(p\cdot \bar{k}+\bar{k}\cdot k_++p\cdot k_++M_K^2\right)}$
& 
$\frac{-3 p.\bar{k}-2 \bar{k}.k_+-p.k_+-2 M_K^2}{2  F_{\pi }
   \left(p.\bar{k}+\bar{k}.k_++p.k_++M_K^2\right)}$ \\
\hline
\end{tabular}
\end{center}
\end{table*}

\newpage

\subsubsection{$J_\mu(\eta_8(\eta) \overline{K}^0(\bar k) {K}^+(k_+))$}

\begin{table*}[!ht]
\begin{center}
\caption{Comparison of the coefficients of external hadron momentum
from Ref.~\cite{Aubrecht:1981cr} and from AMPC for the hadron
current $J_\mu(\eta_8(\eta) \overline{K}^0(\bar k) {K}^+(k_+))$}
 \begin{tabular}{ l || c | c }
\hline     
  Coefficients & Ref.~\cite{Aubrecht:1981cr} & AMPC\\ 
\hline                
  $\eta^\mu$ 
& 
$\frac{3 \eta \cdot \bar{k}+6 \bar{k}\cdot k_++3 \eta \cdot k_++6 M_K^2-2 M_{\pi }^2}
{\sqrt{3} F_{\pi } \left(2 \eta \cdot \bar{k}+2
   \bar{k}\cdot k_++2 \eta \cdot k_++2 M_K^2+M_{\eta }^2-M_{\pi }^2\right)}$ 
& 
$\frac{9 \eta .\bar{k}+18 \bar{k}.k_++9 \eta .k_++14 M_K^2+3 M_{\eta }^2-5 M_{\pi }^2}
{3 \sqrt{3} F_{\pi } \left(2 \eta .\bar{k}+2
   \bar{k}.k_++2 \eta .k_++2 M_K^2+M_{\eta }^2-M_{\pi }^2\right)}$ \\
  $\bar k^\mu$ 
& 
$\frac{-3 \eta \cdot \bar{k}-3 \eta \cdot k_+-3 M_{\eta }^2+M_{\pi
   }^2}{\sqrt{3} F_{\pi } \left(2 \eta \cdot \bar{k}+2 \bar{k}\cdot k_++2 
\eta \cdot k_++2 M_K^2+M_{\eta }^2-M_{\pi }^2\right)}$
& 
$\frac{-9 \eta .\bar{k}-9 \eta .k_+-4 M_K^2-6 M_{\eta }^2+4 M_{\pi }^2}{3 \sqrt{3}
   F_{\pi } \left(2 \eta .\bar{k}+2 \bar{k}.k_++2 \eta .k_++2 M_K^2+M_{\eta }^2-M_{\pi }^2\right)}$ \\
  $k_+^\mu$ 
& 
$\frac{-3 \eta \cdot
   \bar{k}-3 \eta \cdot k_+-3 M_{\eta }^2+M_{\pi }^2}{\sqrt{3} F_{\pi } 
\left(2 \eta \cdot \bar{k}+2 \bar{k}\cdot k_++2 \eta \cdot k_++2
   M_K^2+M_{\eta }^2-M_{\pi }^2\right)}$
& 
$\frac{-9 \eta .\bar{k}-9 \eta .k_+-4 M_K^2-6
   M_{\eta }^2+4 M_{\pi }^2}{3 \sqrt{3} F_{\pi } \left(2 \eta .\bar{k}+2 \bar{k}.k_++2 
\eta .k_++2 M_K^2+M_{\eta }^2-M_{\pi }^2\right)}$ \\
\hline
\end{tabular}
\end{center}
\end{table*}

In the above table the numerator of each expression is further simplified using
GMO relation, and presented below.

\begin{table*}[!ht]
\begin{center}
\caption{Comparison of the coefficients of external hadron momentum
from Ref.~\cite{Aubrecht:1981cr} and from AMPC for the hadron
current $J_\mu(\eta_8(\eta) \overline{K}^0(\bar k) {K}^+(k_+))$}
 \begin{tabular}{ l || c | c }
\hline     
  Coefficients & Ref.~\cite{Aubrecht:1981cr} & AMPC\\ 
\hline                
  $\eta^\mu$ 
& 
$\frac{3 \eta \cdot \bar{k}+6 \bar{k}\cdot k_++3 \eta \cdot k_++6 M_K^2-2 M_{\pi }^2}
{\sqrt{3} F_{\pi } \left(2 \eta \cdot \bar{k}+2
   \bar{k}\cdot k_++2 \eta \cdot k_++2 M_K^2+M_{\eta }^2-M_{\pi }^2\right)}$ 
& 
$\frac{9 \eta .\bar{k}+18 \bar{k}.k_++9 \eta .k_++18 M_K^2-6 M_{\pi }^2}
{3 \sqrt{3} F_{\pi } \left(2 \eta .\bar{k}+2
   \bar{k}.k_++2 \eta .k_++2 M_K^2+M_{\eta }^2-M_{\pi }^2\right)}$ \\
  $\bar k^\mu$ 
& 
$\frac{-3 \eta \cdot \bar{k}-3 \eta \cdot k_+-3 M_{\eta }^2+M_{\pi
   }^2}{\sqrt{3} F_{\pi } \left(2 \eta \cdot \bar{k}+2 \bar{k}\cdot k_++2 
\eta \cdot k_++2 M_K^2+M_{\eta }^2-M_{\pi }^2\right)}$
& 
$\frac{-9 \eta .\bar{k}-9 \eta .k_+-12 M_K^2+6 M_{\pi }^2}{3 \sqrt{3}
   F_{\pi } \left(2 \eta .\bar{k}+2 \bar{k}.k_++2 \eta .k_++2 M_K^2+M_{\eta }^2-M_{\pi }^2\right)}$ \\
  $k_+^\mu$ 
& 
$\frac{-3 \eta \cdot
   \bar{k}-3 \eta \cdot k_+-3 M_{\eta }^2+M_{\pi }^2}{\sqrt{3} F_{\pi } 
\left(2 \eta \cdot \bar{k}+2 \bar{k}\cdot k_++2 \eta \cdot k_++2
   M_K^2+M_{\eta }^2-M_{\pi }^2\right)}$
& 
$\frac{-9 \eta .\bar{k}-9 \eta .k_+-12 M_K^2+6 M_{\pi }^2}
{3 \sqrt{3} F_{\pi } \left(2 \eta .\bar{k}+2 \bar{k}.k_++2 
\eta .k_++2 M_K^2+M_{\eta }^2-M_{\pi }^2\right)}$ \\
\hline
\end{tabular}
\end{center}
\end{table*}

\subsubsection{$J_\mu(K^+_1(k_1)K^+_2(k_2) K^-(k_-))$}

\begin{table*}[!ht]
\begin{center}
\caption{Comparison of the coefficients of external hadron momentum
from Ref.~\cite{Aubrecht:1981cr} and from AMPC for the hadron
current $J_\mu(K^+_1(k_1)K^+_2(k_2) K^-(k_-))$}
 \begin{tabular}{ l || c | c }
\hline     
  Coefficients & Ref.~\cite{Aubrecht:1981cr} & AMPC\\ 
\hline                
  $k_1^\mu$ 
& 
$\frac{\sqrt{2} \left(3 k_-\cdot k_1+3 k_-\cdot k_2+2 M_K^2\right)}
{3 F_{\pi } \left(k_-\cdot k_1+k_-\cdot k_2+k_1\cdot
   k_2+M_K^2\right)}$ 
& 
$-\frac{\sqrt{2}(-k_-.k_1-k_-.k_2-M_K^2)}{F_{\pi } 
\left(k_-.k_1+k_-.k_2+k_1.k_2+M_K^2\right)}$ \\
  $k_2^\mu$ 
& 
$\frac{\sqrt{2} \left(3 k_-\cdot k_1+3 k_-\cdot k_2+2 M_K^2\right)}
{3 F_{\pi } \left(k_-\cdot k_1+k_-\cdot k_2+k_1\cdot
   k_2+M_K^2\right)}$
& 
$-\frac{\sqrt{2}(-k_-.k_1-k_-.k_2-M_K^2)}{F_{\pi }
   \left(k_-.k_1+k_-.k_2+k_1.k_2+M_K^2\right)}$ \\
  $k_-^\mu$ 
& 
$\frac{\sqrt{2} \left(-3 k_-\cdot k_1-3 k_-\cdot k_2-6 k_1\cdot k_2-4 M_K^2\right)}
{3 F_{\pi } \left(k_-\cdot k_1+k_-\cdot
   k_2+k_1\cdot k_2+M_K^2\right)}$
& 
$-\frac{\sqrt{2}(k_-.k_1+k_-.k_2+2 k_1.k_2+M_K^2)}{F_{\pi } \left(k_-.k_1+k_-.k_2+k_1.k_2+M_K^2\right)}$ \\
\hline
\end{tabular}
\end{center}
\end{table*}

\subsubsection{$J_\mu(K^0(k)\overline{K}^0(\bar k) K^+(k_+))$}

\begin{table*}[!ht]
\begin{center}
\caption{Comparison of the coefficients of external hadron momentum
from Ref.~\cite{Aubrecht:1981cr} and from AMPC for the hadron
current $J_\mu(K^0(k)\overline{K}^0(\bar k) K^+(k_+))$}
 \begin{tabular}{ l || c | c }
\hline     
  Coefficients & Ref.~\cite{Aubrecht:1981cr} & AMPC\\ 
\hline                
  $k^\mu$ 
& 
$\frac{3 k\cdot \bar{k}+3 \bar{k}\cdot k_++2 M_K^2}
{3 \sqrt{2} F_{\pi } \left(k\cdot \bar{k}+\bar{k}\cdot k_++k\cdot
   k_++M_K^2\right)}$ 
& 
$\frac{k.\bar{k}+\bar{k}.k_++M_K^2}{\sqrt{2} F_{\pi } 
\left(k.\bar{k}+\bar{k}.k_++k.k_++M_K^2\right)}$ \\
  $\bar k^\mu$ 
& 
$\frac{-3 k\cdot \bar{k}-3 \bar{k}\cdot k_+-6 k\cdot k_+-4 M_K^2}
{3 \sqrt{2} F_{\pi } \left(k\cdot \bar{k}+\bar{k}\cdot
   k_++k\cdot k_++M_K^2\right)}$
& 
$\frac{-k.\bar{k}-\bar{k}.k_+-2 k.k_+-M_K^2}{\sqrt{2}
   F_{\pi } \left(k.\bar{k}+\bar{k}.k_++k.k_++M_K^2\right)}$ \\
  $k_+^\mu$ 
& 
$\frac{3 k\cdot \bar{k}+3 \bar{k}\cdot k_++2 M_K^2}
{3 \sqrt{2} F_{\pi } \left(k\cdot \bar{k}+\bar{k}\cdot k_++k\cdot
   k_++M_K^2\right)}$
& 
$\frac{k.\bar{k}+\bar{k}.k_++M_K^2}{\sqrt{2} F_{\pi }
   \left(k.\bar{k}+\bar{k}.k_++k.k_++M_K^2\right)}$ \\
\hline
\end{tabular}
\end{center}
\end{table*}

\subsubsection{$J_\mu(\pi^0(p_1)\pi^0(p_2) K^+(k_+))$}

\begin{table*}[!ht]
\begin{center}
\caption{Comparison of the coefficients of external hadron momentum
from Ref.~\cite{Aubrecht:1981cr} and from AMPC for the hadron
current $J_\mu(\pi^0(p_1)\pi^0(p_2) K^+(k_+))$}
 \begin{tabular}{ l || c | c }
\hline     
  Coefficients & Ref.~\cite{Aubrecht:1981cr} & AMPC\\ 
\hline                
  $p_1^\mu$ 
& 
$\frac{-3 k_+\cdot p_1-3 k_+\cdot p_2-2 M_K^2}
{6 \sqrt{2} F_{\pi } \left(k_+\cdot p_1+k_+
\cdot p_2+M_{\pi }^2+p_1\cdot p_2\right)}$ 
& 
$\frac{-k_+.p_1-k_+.p_2}{2\sqrt{2} F_{\pi } 
\left(k_+.p_1+k_+.p_2+M_{\pi }^2+p_1.p_2\right)}$ \\
  $p_2^\mu$ 
& 
$\frac{-3
   k_+\cdot p_1-3 k_+\cdot p_2-2 M_K^2}{6 \sqrt{2} F_{\pi } 
\left(k_+\cdot p_1+k_+\cdot p_2+M_{\pi }^2+p_1\cdot p_2\right)}$
& 
$\frac{-k_+.p_1-k_+.p_2}{2\sqrt{2} F_{\pi }
   \left(k_+.p_1+k_+.p_2+M_{\pi }^2+p_1.p_2\right)}$ \\
  $k_+^\mu$ 
& 
$\frac{3 k_+\cdot p_1+3
   k_+\cdot p_2-2 M_K^2+6 M_{\pi }^2+6 p_1\cdot p_2}
{6 \sqrt{2} F_{\pi } \left(k_+\cdot p_1+k_+
\cdot p_2+M_{\pi }^2+p_1\cdot p_2\right)}$
& 
$\frac{k_+.p_1+k_+.p_2+2 M_{\pi }^2+2 p_1.p_2}
{2\sqrt{2} F_{\pi } \left(k_+.p_1+k_+.p_2+M_{\pi
   }^2+p_1.p_2\right)}$ \\
\hline
\end{tabular}
\end{center}
\end{table*}

\subsubsection{$J_\mu(\pi^+(p_+)\pi^-(p_-) K^+(k_+))$}

\begin{table*}[!ht]
\begin{center}
\caption{Comparison of the coefficients of external hadron momentum
from Ref.~\cite{Aubrecht:1981cr} and from AMPC for the hadron
current $J_\mu(\pi^+(p_+)\pi^-(p_-) K^+(k_+))$}
 \begin{tabular}{ l || c | c }
\hline     
  Coefficients & Ref.~\cite{Aubrecht:1981cr} & AMPC\\ 
\hline                
  $p_+^\mu$ 
& 
$\frac{3 p_-\cdot k_+-M_K^2+3 M_{\pi }^2+3 p_-\cdot p_+}
{3 \sqrt{2} F_{\pi } \left(p_-\cdot k_++k_+\cdot p_++M_{\pi }^2+p_-\cdot
   p_+\right)}$ 
& 
$-\frac{-p_-.k_+-M_{\pi }^2-p_-.p_+}
{\sqrt{2} F_{\pi } \left(p_-.k_++k_+.p_++M_{\pi }^2+p_-.p_+\right)}$ \\
  $p_-^\mu$ 
& 
$\frac{-3 p_-\cdot k_+-6 k_+\cdot p_+-M_K^2-3 M_{\pi }^2-3 p_-\cdot p_+}
{3 \sqrt{2} F_{\pi } \left(p_-\cdot k_++k_+\cdot p_++M_{\pi
   }^2+p_-\cdot p_+\right)}$
& 
$-\frac{p_-.k_++2 k_+.p_++M_{\pi
   }^2+p_-.p_+}{\sqrt{2} F_{\pi } \left(p_-.k_++k_+.p_++M_{\pi }^2+p_-.p_+\right)}$ \\
  $k_+^\mu$ 
& 
$\frac{3 p_-\cdot k_+-M_K^2+3 M_{\pi }^2+3 p_-\cdot p_+}
{3 \sqrt{2} F_{\pi } \left(p_-\cdot k_++k_+\cdot p_++M_{\pi
   }^2+p_-\cdot p_+\right)}$
& 
$-\frac{-p_-.k_+-M_{\pi }^2-p_-.p_+}{\sqrt{2} F_{\pi }
   \left(p_-.k_++k_+.p_++M_{\pi }^2+p_-.p_+\right)}$ \\
\hline
\end{tabular}
\end{center}
\end{table*}

\newpage

\subsubsection{$J_\mu(\pi^0(p)\pi^+(p_+) K^0(k_0))$}

\begin{table*}[!ht]
\begin{center}
\caption{Comparison of the coefficients of external hadron momentum
from Ref.~\cite{Aubrecht:1981cr} and from AMPC for the hadron
current $J_\mu(\pi^0(p)\pi^+(p_+) K^0(k_0))$}
 \begin{tabular}{ l || c | c }
\hline     
  Coefficients & Ref.~\cite{Aubrecht:1981cr} & AMPC\\ 
\hline                
  $p_+^\mu$ 
& 
$\frac{-3 p\cdot k_0-p_+\cdot k_0-2 M_{\pi }^2-2 p\cdot p_+}
{2 F_{\pi } \left(p\cdot k_0+p_+\cdot k_0+M_{\pi }^2+p\cdot
   p_+\right)}$ 
& 
$\frac{-3 p.k_0-p_+.k_0-2 M_{\pi }^2-2 p.p_+}
{2  F_{\pi } \left(p.k_0+p_+.k_0+M_{\pi }^2+p.p_+\right)}$ \\
  $p^\mu$ 
& 
$\frac{p\cdot k_0+3 p_+\cdot k_0+2 M_{\pi }^2+2 p\cdot p_+}
{2 F_{\pi } \left(p\cdot k_0+p_+\cdot k_0+M_{\pi }^2+p\cdot
   p_+\right)}$
& 
$\frac{p.k_0+3 p_+.k_0+2 M_{\pi
   }^2+2 p.p_+}{2  F_{\pi } \left(p.k_0+p_+.k_0+M_{\pi }^2+p.p_+\right)}$ \\
  $k_0^\mu$ 
& 
$\frac{p_+\cdot k_0-p\cdot k_0}{2 F_{\pi } 
\left(p\cdot k_0+p_+\cdot k_0+M_{\pi }^2+p\cdot p_+\right)}$
& 
$\frac{p_+.k_0-p.k_0}{2  F_{\pi }
   \left(p.k_0+p_+.k_0+M_{\pi }^2+p.p_+\right)}$ \\
\hline
\end{tabular}
\end{center}
\end{table*}

\subsubsection{$J_\mu(\eta_8(\eta)\pi^+(p_+) K^0(k))$}

\begin{table*}[!ht]
\begin{center}
\caption{Comparison of the coefficients of external hadron momentum
from Ref.~\cite{Aubrecht:1981cr} and from AMPC for the hadron
current $J_\mu(\eta_8(\eta)\pi^+(p_+) K^0(k))$}
 \begin{tabular}{ l || c | c }
\hline     
  Coefficients & Ref.~\cite{Aubrecht:1981cr} & AMPC\\ 
\hline                
  $\eta^\mu$ 
& 
$\frac{-3 k\cdot p_+-3 k\cdot \eta -2 M_K^2}
{\sqrt{3} F_{\pi } \left(2 k\cdot p_++2 k\cdot \eta +M_{\eta }^2+M_{\pi }^2+2 \eta \cdot
   p_+\right)}$ 
& 
$\frac{-18 k.p_+-18 k.\eta -8 M_K^2-3 M_{\eta }^2-M_{\pi }^2}
{6 \sqrt{3} F_{\pi } \left(2 k.p_++2 k.\eta +M_{\eta }^2+M_{\pi }^2+2 \eta
   .p_+\right)}$ \\
  $p_+^\mu$ 
& 
$\frac{-3 k\cdot p_+-3 k\cdot \eta -2 M_K^2}
{\sqrt{3} F_{\pi } \left(2 k\cdot p_++2 k\cdot \eta +M_{\eta }^2+M_{\pi }^2+2 \eta \cdot
   p_+\right)}$
& 
$\frac{-18 k.p_+-18 k.\eta -8 M_K^2-3 M_{\eta }^2-M_{\pi }^2}
{6 \sqrt{3} F_{\pi } \left(2 k.p_++2 k.\eta +M_{\eta }^2+M_{\pi }^2+2
   \eta .p_+\right)}$ \\
  $k^\mu$ 
& 
$\frac{3 k\cdot p_++3 k\cdot \eta -2 M_K^2+3 M_{\eta }^2
+3 M_{\pi }^2+6 \eta \cdot p_+}{\sqrt{3} F_{\pi } \left(2 k\cdot p_++2 k\cdot
   \eta +M_{\eta }^2+M_{\pi }^2+2 \eta \cdot p_+\right)}$
& 
$\frac{18 k.p_++18 k.\eta -8 M_K^2+15 M_{\eta }^2+17 
M_{\pi }^2+36 \eta .p_+}{6 \sqrt{3} F_{\pi } \left(2 k.p_++2 k.\eta
   +M_{\eta }^2+M_{\pi }^2+2 \eta .p_+\right)}$ \\
\hline
\end{tabular}
\end{center}
\end{table*}

In the above table the numerator of each expression is further simplified using
GMO relation, and presented below.

\begin{table*}[!ht]
\begin{center}
\caption{Comparison of the coefficients of external hadron momentum
from Ref.~\cite{Aubrecht:1981cr} and from AMPC for the hadron
current $J_\mu(\eta_8(\eta)\pi^+(p_+) K^0(k))$}
 \begin{tabular}{ l || c | c }
\hline     
  Coefficients & Ref.~\cite{Aubrecht:1981cr} & AMPC\\ 
\hline                
  $\eta^\mu$ 
& 
$\frac{-3 k\cdot p_+-3 k\cdot \eta -2 M_K^2}
{\sqrt{3} F_{\pi } \left(2 k\cdot p_++2 k\cdot \eta +M_{\eta }^2+M_{\pi }^2+2 \eta \cdot
   p_+\right)}$ 
& 
$\frac{-18 k.p_+-18 k.\eta -12 M_K^2}
{6 \sqrt{3} F_{\pi } \left(2 k.p_++2 k.\eta +M_{\eta }^2+M_{\pi }^2+2 \eta
   .p_+\right)}$ \\
  $p_+^\mu$ 
& 
$\frac{-3 k\cdot p_+-3 k\cdot \eta -2 M_K^2}
{\sqrt{3} F_{\pi } \left(2 k\cdot p_++2 k\cdot \eta +M_{\eta }^2+M_{\pi }^2+2 \eta \cdot
   p_+\right)}$
& 
$\frac{-18 k.p_+-18 k.\eta -12 M_K^2}
{6 \sqrt{3} F_{\pi } \left(2 k.p_++2 k.\eta +M_{\eta }^2+M_{\pi }^2+2
   \eta .p_+\right)}$ \\
  $k^\mu$ 
& 
$\frac{3 k\cdot p_++3 k\cdot \eta +2 M_K^2
+2 M_{\pi }^2+6 \eta \cdot p_+}{\sqrt{3} F_{\pi } \left(2 k\cdot p_++2 k\cdot
   \eta +M_{\eta }^2+M_{\pi }^2+2 \eta \cdot p_+\right)}$
& 
$\frac{18 k.p_++18 k.\eta +12 M_K^2+12 
M_{\pi }^2+36 \eta .p_+}{6 \sqrt{3} F_{\pi } \left(2 k.p_++2 k.\eta
   +M_{\eta }^2+M_{\pi }^2+2 \eta .p_+\right)}$ \\
\hline
\end{tabular}
\end{center}
\end{table*}

\newpage

\subsubsection{$J_\mu(\eta_8(\eta)\pi^0(p) K^+(k_+))$}

\begin{table*}[!ht]
\begin{center}
\caption{Comparison of the coefficients of external hadron momentum
from Ref.~\cite{Aubrecht:1981cr} and from AMPC for the hadron
current $J_\mu(\eta_8(\eta)\pi^0(p) K^+(k_+))$}
 \begin{tabular}{ l || c | c }
\hline     
  Coefficients & Ref.~\cite{Aubrecht:1981cr} & AMPC\\ 
\hline                
  $\eta^\mu$ 
& 
$\frac{-3 p\cdot k_+-3 \eta \cdot k_+-2 M_K^2}
{\sqrt{6} F_{\pi } \left(2 p\cdot k_++2 \eta \cdot k_++M_{\eta }^2+M_{\pi }^2+2 p\cdot \eta
   \right)}$ 
& 
$\frac{-18 p.k_+-18 \eta .k_+-8 M_K^2-3 M_{\eta }^2-M_{\pi }^2}
{6 \sqrt{6} F_{\pi } \left(2 p.k_++2 \eta .k_++M_{\eta }^2+M_{\pi }^2+2
   p.\eta \right)}$ \\
  $p^\mu$ 
& 
$\frac{-3 p\cdot k_+-3 \eta \cdot k_+-2 M_K^2}
{\sqrt{6} F_{\pi } \left(2 p\cdot k_++2 \eta \cdot k_++M_{\eta }^2+M_{\pi }^2+2 p\cdot
   \eta \right)}$
& 
$\frac{-18 p.k_+-18 \eta .k_+-8 M_K^2-3 M_{\eta }^2-M_{\pi }^2}
{6 \sqrt{6} F_{\pi } \left(2 p.k_++2 \eta .k_++M_{\eta }^2+M_{\pi
   }^2+2 p.\eta \right)}$ \\
  $k_+^\mu$ 
& 
$\frac{3 p\cdot k_++3 \eta \cdot k_+-2 M_K^2+3 M_{\eta }^2+
3 M_{\pi }^2+6 p\cdot \eta }{\sqrt{6} F_{\pi } \left(2 p\cdot k_++2 \eta
   \cdot k_++M_{\eta }^2+M_{\pi }^2+2 p\cdot \eta \right)}$
& 
$\frac{18 p.k_++18 \eta .k_+-8 M_K^2+15 M_{\eta }^2+17 M_{\pi }^2+36 p.\eta }{6 \sqrt{6} F_{\pi } \left(2 p.k_++2 \eta
   .k_++M_{\eta }^2+M_{\pi }^2+2 p.\eta \right)}$ \\
\hline
\end{tabular}
\end{center}
\end{table*}

In the above table the numerator of each expression is further simplified using
GMO relation, and presented below.

\begin{table*}[!ht]
\begin{center}
\caption{Comparison of the coefficients of external hadron momentum
from Ref.~\cite{Aubrecht:1981cr} and from AMPC for the hadron
current $J_\mu(\eta_8(\eta)\pi^0(p) K^+(k_+))$}
 \begin{tabular}{ l || c | c }
\hline     
  Coefficients & Ref.~\cite{Aubrecht:1981cr} & AMPC\\ 
\hline                
  $\eta^\mu$ 
& 
$\frac{-3 p\cdot k_+-3 \eta \cdot k_+-2 M_K^2}
{\sqrt{6} F_{\pi } \left(2 p\cdot k_++2 \eta \cdot k_++M_{\eta }^2+M_{\pi }^2+2 p\cdot \eta
   \right)}$ 
& 
$\frac{-18 p.k_+-18 \eta .k_+-12 M_K^2}
{6 \sqrt{6} F_{\pi } \left(2 p.k_++2 \eta .k_++M_{\eta }^2+M_{\pi }^2+2
   p.\eta \right)}$ \\
  $p^\mu$ 
& 
$\frac{-3 p\cdot k_+-3 \eta \cdot k_+-2 M_K^2}
{\sqrt{6} F_{\pi } \left(2 p\cdot k_++2 \eta \cdot k_++M_{\eta }^2+M_{\pi }^2+2 p\cdot
   \eta \right)}$
& 
$\frac{-18 p.k_+-18 \eta .k_+-12 M_K^2}
{6 \sqrt{6} F_{\pi } \left(2 p.k_++2 \eta .k_++M_{\eta }^2+M_{\pi
   }^2+2 p.\eta \right)}$ \\
  $k_+^\mu$ 
& 
$\frac{3 p\cdot k_++3 \eta \cdot k_++2 M_K^2+
2 M_{\pi }^2+6 p\cdot \eta }{\sqrt{6} F_{\pi } \left(2 p\cdot k_++2 \eta
   \cdot k_++M_{\eta }^2+M_{\pi }^2+2 p\cdot \eta \right)}$
& 
$\frac{18 p.k_++18 \eta .k_++12 M_K^2+12 M_{\pi }^2+36 p.\eta }{6 \sqrt{6} F_{\pi } \left(2 p.k_++2 \eta
   .k_++M_{\eta }^2+M_{\pi }^2+2 p.\eta \right)}$ \\
\hline
\end{tabular}
\end{center}
\end{table*}

\subsubsection{$J_\mu(\eta_8(\eta_1)\eta_8(\eta_2) K^+(k_+))$}

\begin{table*}[!ht]
\begin{center}
\caption{Comparison of the coefficients of external hadron momentum
from Ref.~\cite{Aubrecht:1981cr} and from AMPC for the hadron
current $J_\mu(\eta_8(\eta_1)\eta_8(\eta_2) K^+(k_+))$}
 \begin{tabular}{ l || c | c }
\hline     
  Coefficients & Ref.~\cite{Aubrecht:1981cr} & AMPC\\ 
\hline                
  $\eta_1^\mu$ 
& 
$\frac{-3 k_+\cdot \eta _1-3 k_+\cdot \eta _2-2 M_K^2}
{2 \sqrt{2} F_{\pi } \left(k_+\cdot \eta _1+k_+\cdot \eta _2+M_{\eta }^2+\eta _1\cdot
   \eta _2\right)}$ 
& 
$\frac{-9 k_+.\eta _1-9 k_+.\eta _2-3 M_{\eta }^2-M_{\pi }^2}
{6\sqrt{2} F_{\pi } \left(k_+.\eta _1+k_+.\eta _2+M_{\eta }^2+\eta _1.\eta
   _2\right)}$ \\
  $\eta_2^\mu$ 
& 
$\frac{-3 k_+\cdot \eta _1-3 k_+\cdot \eta _2-2 M_K^2}
{2 \sqrt{2} F_{\pi } \left(k_+\cdot \eta _1+k_+\cdot \eta _2+M_{\eta
   }^2+\eta _1\cdot \eta _2\right)}$
& 
$\frac{-9 k_+.\eta _1-9 k_+.\eta _2-3 M_{\eta }^2-M_{\pi }^2}
{6\sqrt{2} F_{\pi } \left(k_+.\eta _1+k_+.\eta _2+M_{\eta }^2+\eta _1.\eta
   _2\right)}$ \\
  $k_+^\mu$ 
& 
$\frac{3 k_+\cdot \eta _1+3 k_+\cdot \eta _2-2 M_K^2+6 M_{\eta }^2+6 \eta _1\cdot \eta _2}
{2 \sqrt{2} F_{\pi }
   \left(k_+\cdot \eta _1+k_+\cdot \eta _2+M_{\eta }^2+\eta _1\cdot \eta _2\right)}$
& 
$\frac{9 k_+.\eta _1+9 k_+.\eta _2+15 M_{\eta }^2-
M_{\pi }^2+18 \eta _1.\eta _2}{6\sqrt{2} F_{\pi } \left(k_+.\eta _1+k_+.\eta _2+M_{\eta
   }^2+\eta _1.\eta _2\right)}$ \\
\hline
\end{tabular}
\end{center}
\end{table*}

In the above table the numerator of each expression is further simplified using
GMO relation, and presented below.

\begin{table*}[!ht]
\begin{center}
\caption{Comparison of the coefficients of external hadron momentum
from Ref.~\cite{Aubrecht:1981cr} and from AMPC for the hadronic
current $J_\mu(\eta_8(\eta_1)\eta_8(\eta_2) K^+(k_+))$}
 \begin{tabular}{ l || c | c }
\hline     
  Coefficients & Ref.~\cite{Aubrecht:1981cr} & AMPC\\ 
\hline                
  $\eta_1^\mu$ 
& 
$\frac{-3 k_+\cdot \eta _1-3 k_+\cdot \eta _2-2 M_K^2}
{2 \sqrt{2} F_{\pi } \left(k_+\cdot \eta _1+k_+\cdot \eta _2+M_{\eta }^2+\eta _1\cdot
   \eta _2\right)}$ 
& 
$\frac{-9 k_+.\eta _1-9 k_+.\eta _2-4 M_{K }^2}
{6\sqrt{2} F_{\pi } \left(k_+.\eta _1+k_+.\eta _2+M_{\eta }^2+\eta _1.\eta
   _2\right)}$ \\
  $\eta_2^\mu$ 
& 
$\frac{-3 k_+\cdot \eta _1-3 k_+\cdot \eta _2-2 M_K^2}
{2 \sqrt{2} F_{\pi } \left(k_+\cdot \eta _1+k_+\cdot \eta _2+M_{\eta
   }^2+\eta _1\cdot \eta _2\right)}$
& 
$\frac{-9 k_+.\eta _1-9 k_+.\eta _2-4 M_{K}^2}
{6\sqrt{2} F_{\pi } \left(k_+.\eta _1+k_+.\eta _2+M_{\eta }^2+\eta _1.\eta
   _2\right)}$ \\
  $k_+^\mu$ 
& 
$\frac{3 k_+\cdot \eta _1+3 k_+\cdot \eta _2+6 M_K^2-2 M_{\pi }^2+6 \eta _1\cdot \eta _2}
{2 \sqrt{2} F_{\pi }
   \left(k_+\cdot \eta _1+k_+\cdot \eta _2+M_{\eta }^2+\eta _1\cdot \eta _2\right)}$
& 
$\frac{9 k_+.\eta _1+9 k_+.\eta _2+20 M_{K }^2-
6 M_{\pi }^2+18 \eta _1.\eta _2}{6\sqrt{2} F_{\pi } \left(k_+.\eta _1+k_+.\eta _2+M_{\eta
   }^2+\eta _1.\eta _2\right)}$ \\
\hline
\end{tabular}
\end{center}
\end{table*}


\section{Summary \label{sec:summary}}
In this report, we have presented results of our checks of the consistency of AMPC and established results. In the 
meson sector, we have analyzed accessible form factors and scattering amplitudes including the 
kaon-Compton process. As long as the number of particles is manageable, explicit checks were tractable. Large
number of notebooks are provided. This work was spurned by our recent investigations of the $K \rightarrow \pi l^+l^-$
process where we discovered that the $G_{27}$ piece was not published in the literature.
Since AMPC is very versatile, we have used it in the non-leptonic kaon decay sector to isolate 
the contributions of the odd-intrinsic parity sector that also involves $G_{27}$ and the higher order 
pieces as well. All the details of the notations are given in \cite{Unterdorfer:2005au}.
Another application is to the tree-level chiral processes appearing in $\tau$-decays. As recently as \cite{Usuki:2009zza}, BELLE
was using the results of \cite{Aubrecht:1981cr} which neglected the quark masses. We give the quark mass corrected
results here. It is our belief that AMPC can be used to obtain amplitudes such as $\pi\gamma \rightarrow \pi\pi$ and others of
importance to the COMPASS experiment as well. By providing explicit notebooks, we believe we have provided a
service to the community which can also be used as a learning aid.

\vskip0.5cm
{\small {\bf Acknowledgements:} 
We are very grateful to Gerhard Ecker for extensive correspondence on the subject.  We also thank J. R. Pelaez for a discussion. 
BA and ISI thank Rahul Sinha and Shrihari Gopalakrishna for their hospitality
at IMSc when part of this work was done. BA and ISI would also like to thank Rajita Menon for
discussions at an early stage of this work. DD would like to thank Gauhar Abbas,
Monalisa Patra and the Centre for High Energy Physics for hospitality during early stage of the work.}

\end{document}